\documentclass[usenatbib,natbib2,natbibmnfix]{mn2e}
\usepackage{mathptm}
\usepackage{times}

\usepackage{epsfig}

\newcommand{\kpch}{$h^{-1} \rm kpc$~}

\newcommand{\Mo}{\rm M_\odot}
\newcommand{\LCDM}{$\Lambda$CDM}

\title[]
{Where will supersymmetric dark matter first be seen?}
\author[Gao et al.]
       {L. Gao,$^{1,2}$\thanks{Email:lgao@bao.ac.cn},
         C. S. Frenk$^2$,
          A. Jenkins$^2$, 
          V. Springel$^{3,4}$,
         S. D. M. White$^5$
\\
$^1$Partner Group of the Max Planck Institute for Astrophysics,
National Astronomical Observatories, Chinese Academy of Sciences,
Beijing, 100012, China \\
$^2$Institute of Computational Cosmology, Department of Physics,
University of Durham,\\ Science Laboratories, South Road, Durham DH1
3LE \\
$^3$Heidelberger Institut f\"{u}r Theoretische Studien, Schloss-Wolfsbrunnenweg 35, 69118 Heidelberg, Germany\\
$^4$Zentrum f\"ur Astronomie der Universit\"at Heidelberg, Astronomisches
Recheninstitut, M\"{o}nchhofstr. 12-14, 69120 Heidelberg, Germany\\
$^5$Max-Planck Institute for Astrophysics, Karl-Schwarzschild Str. 1,
D-85748, Garching, Germany \\
}

\begin{document}
\label{firstpage} \maketitle
\begin{abstract}
If the dark matter consists of supersymmetric particles, $\gamma$-ray
observatories such as the Large Area Telescope aboard the Fermi
satellite may detect annihilation radiation from the haloes of
galaxies and galaxy clusters. Much recent effort has been devoted to
searching for this signal around the Milky Way's dwarf
satellites. Using a new suite of high-resolution simulations of galaxy
cluster haloes (the Phoenix Project), together with the Aquarius
simulations of Milky-Way-like galaxy haloes, we show that higher
signal-to-noise and equally clean signals are, in fact, predicted to
come from nearby rich galaxy clusters. Most of the cluster emission is
produced by small subhaloes with masses less than that of the Sun. The
large range of mass scales covered by our two sets of simulations
allows us to deduce a physically motivated extrapolation to these
small (and unresolved) masses. Since tidal effects destroy subhaloes
in the dense inner regions of haloes, most cluster emission is then
predicted to come from large radii, implying that the nearest and
brightest systems should be much more extended than Fermi's angular
resolution limit. The most promising targets for detection are
clusters such as Coma and Fornax, but detection algorithms must be
tuned to the predicted profile of the emission if they are to maximize
the chance of finding this weak signal.
\end{abstract}

\title{Dark matter}

\section{Introduction}

Annihilation radiation at $\gamma$-ray frequencies offers one of the
most exciting prospects for non-gravitational detection of cold dark
matter, and is expected if the dark matter consists of supersymmetric
particles~\citep[e.g.][]{Berezinsky1994,Berezinsky2003,Bergstrom1998,Stoehr2003,Koushiappas2004,
Colafrancesco2007,Diemand2007,Kuhlen2008,Pieri2008,sp08a,Strigari2008,jeltema10,ackermann10,Zavala2010}.
Much effort is being devoted to searching for this signal around the
Milky Way's dwarf companions, in particular using the Fermi
satellite~\citep{abdo}.

Predictions for the properties of the annihilation radiation rely on a
detailed understanding of the structure of cold dark matter haloes which can
be gained only through high-resolution numerical simulations of halo
formation.  The structure of galaxy-mass cold dark matter haloes has
been investigated in considerable
depth~\citep[e.g.][]{Diemand2007,diemand08,Kuhlen2008,sp08a,sp08b,Anderson2010,Kamionkowski2010}
showing that the radial distribution of low-mass subhaloes, and thus
of annihilation radiation, is much less centrally concentrated than
that of the dark matter as a whole. In the Milky Way, this results in
the dominant subhalo contribution to the annihilation radiation coming
from large galactrocentric distance and so appearing almost uniform
across the sky to an observer on Earth~\citep{sp08a}. This same effect
causes the annihilation radiation from an external galaxy cluster to
appear much less centrally concentrated than the distribution of
galaxies. As we show below, this has significant implications for the
optimal strategy for detecting the annihilation signal.

In this paper we present some of the largest high-resolution
simulations of cluster haloes to date (the Phoenix Project) and use
them to investigate the detailed structure of the dark matter
distribution in clusters and its halo-to-halo variation. We use these
data, together with data from the Aquarius set of galaxy halo
simulations \citep{sp08b}, to predict the expected $\gamma$-ray
annihilation radiation from cluster haloes which we compare to the
expected annihilation radiation from giant and satellite galaxy
haloes.

As we were completing this work, \cite{Pinzke11} and \cite{sanchez}
posted preprints investigating, amongst other things, the $\gamma$-ray
annihilation radiation expected from galaxy clusters. The luminosity
and spatial distribution of this radiation depend sensitively on the
properties of surviving dark matter subhaloes down to the limiting
mass of the cold dark matter power spectrum, which may be in the range
$10^{-6}$ to $10^{-12}M_\odot$~\citep{Hofmann2001,Green05}.  For their
analysis,
\cite{Pinzke11} relied on an extrapolation of scalings based on published
results for simulations of galactic dark matter haloes, including
those of the Aquarius Project, while \cite{sanchez} extended the
semi-analytic model of \cite{Kamionkowski2010}, rescaling relevant
model parameters. Combining the Phoenix and Aquarius simulations we
test explicitly the validity of the scalings used by
\cite{Pinzke11} and \cite{sanchez}, we investigate their underlying physical basis, and
we thus construct a more robust (though still uncertain) framework for
extrapolation. For the most part, our results are in agreement with
those of \cite{Pinzke11}, but not with those of \cite{sanchez}
who infer a much weaker contribution from subhalos to the total
annihilation radiation from clusters than \cite{Pinzke11} or us find. 
In this study, we also present an estimate of the expected
signal-to-noise of the annihilation radiation from nearby clusters and
compare it to that from nearby dwarf and giant galaxies.

The outline of our paper is as follows. Sections~2 and~3 give brief
descriptions of our simulation suite and of a model for calculating
the annihilation flux and its signal-to-noise in an idealised
experiment. In Section~4, we discuss our results and their
implications for dark matter detection.

\section{Numerical simulations}

The new dark matter simulations analysed in this study come from
the Phoenix Project (Gao et al. 2011, in preparation). We supplement
them with previous high-resolution simulations of galactic halos from
the Aquarius Project carried out by the Virgo Consortium (Springel et
al. 2008a,b). Starting from initial conditions appropriate to the
$\Lambda$CDM cosmology, both sets of simulations integrate the orbits
of large numbers of particles using the Gadget-3 N-body code (see
Springel et al. 2008a).  The cosmological parameters adopted for both
the Aquarius and Phoenix projects are those of Virgo's Millennium
Simulation~\citep{sp05}: $\Omega_m=0.25$, $\Omega_{\Lambda}=0.75$,
$\sigma_8=0.9$, $n_s=1$, and a Hubble constant $H_0=100\,h\,{\rm
km\,s^{-1}}=73\, {\rm km\,s^{-1}Mpc^{-1}}$. These were close to the
best fit values derived from the first year of data from the WMAP
satellite \citep{spergel03} but are not consistent with the parameter
ranges found through analysis of the seven-year WMAP data together with other
large-scale structure observations \citep{Komatsu11}. The small offset
is, however, of no consequence for the topics addressed in this paper.

For the Phoenix Project, we have carried out a suite of extremely high
resolution simulations of the dark matter distribution in galaxy
clusters. This suite consists of nine cluster-size dark matter haloes
with masses in the range $[5-20]\times 10^{14}\,h^{-1}{\rm
M_{\odot}}$. These were selected at random from the Millennium
Simulation and resimulated at various numerical resolutions.  The
largest of these ``Phoenix'' simulations, labelled Ph-A-1, represents
the dark matter with $1.0 \times 10^9$ particles within $r_{200}$, the
radius at which the enclosed mean density is $200$ times the cosmic
critical density. It has a particle mass of $6.4 \times
10^{5}h^{-1}\Mo$ and a Plummer-equivalent force softening of
$0.15$\kpch in comoving coordinates at all times. This particular
cluster has also been simulated at four lower resolution levels
(producing Ph-A-2 to Ph-A-5) in order to assess how resolution affects
inferences about cluster structure. At the next-to-highest resolution
level ($\sim 1.3\times 10^8$ particles within $r_{200}$), we have
simulated an additional eight clusters (Ph-B-2 to Ph-I-2) with a
particle mass of about $5 \times 10^{6}h^{-1}\Mo$ and a force
softening of $0.32$\kpch in order to quantify the cluster-to-cluster
variation in dark matter properties. We will present details of the
Phoenix simulation suite in a forthcoming paper (Gao et al. 2011, in
preparation).

\section{Results}
\subsection{The total cluster surface brightness}
The total $\gamma$-ray annihilation luminosity of a dark matter halo
is the sum of contributions from the smooth main halo, from resolved
subhaloes, and from unresolved subhaloes. (Caustics and tidal streams
make a negligible contribution to the annihilation
luminosity;\cite{Vogelsberger11}.) If the density distribution in the
inner regions of the smooth main halo and the resolved subhaloes are assumed
to be adequately fit by the NFW formula~\citep{NFW96,NFW97} their
emission integrals $\int \rho^2 dV$ can be estimated simply as
$1.23\,V_{\rm max}^4/(G^2r_{\rm max})$ \citep{sp08a}. Here $V_{\rm
  max}$ is the maximum circular velocity of the halo or subhalo and
$r_{\rm max}$ the radius at which this maximum circular velocity is
reached.

\begin{figure}
\begin{center}
\resizebox{8.0cm}{!}{\includegraphics{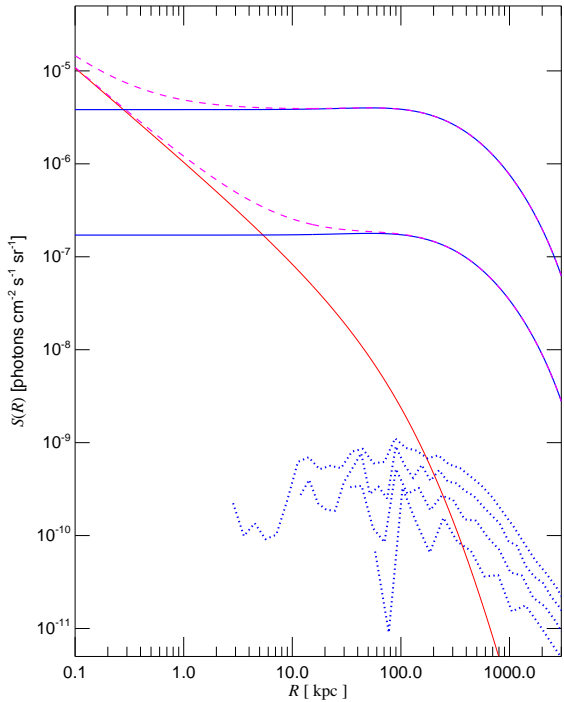}}%
\caption{Surface brightness profiles  from dark matter
annihilation for various components of the Ph-A-1 simulation of a rich
galaxy cluster. Surface brightness is given in units of annihilation
photons per cm$^2$ per second per steradian for fiducial values of
$100\,{\rm Gev}$ for $m_p$, the dark matter particle mass, and $3
\times 10^{-26}{\rm cm}^3 {\rm s}^{-1}$ for $\langle \sigma v\rangle$,
the thermally averaged velocity-weighted annihilation cross-section,
assuming $N_{\gamma}=1$ photons per annihilation. This surface
brightness scales as $N_{\gamma}\langle \sigma
v\rangle/m_p^2$. Projected radius is given in units of kpc. The red
line shows radiation from the smoothly distributed dark matter within
the main component of the cluster. The ragged blue dotted lines
show radiation from resolved dark matter subhaloes with masses
exceeding $5 \times 10^7$, $5 \times 10^8$, $5 \times 10^9$ and $5
\times 10^{10}\,\Mo$ (from top to bottom). Extrapolating to mass
limits of $10^{-6}$ and $10^{-12}\,\Mo$ as discussed in the text gives
rise to the smooth blue curves. The purple dashed lines show the
results of summing smooth and subhalo contributions.}
\label{fig:surface}
\end{center}
\end{figure}

In Figure~\ref{fig:surface}, we show the azimuthally averaged surface
brightness profile for Ph-A-1, split into the components due to the
smooth dark matter distribution and to subhaloes resolved down to four
mass thresholds differing by factors of ten. The subhalo component is
clearly much less centrally concentrated to the cluster centre than
the smooth component and its shape appears independent of mass
threshold as far as can be judged given the noise introduced by the
finite number of subhaloes involved. The overall level of subhalo
emission increases steadily as the threshold decreases. The smallest
subhaloes resolved in Ph-A-1 have masses $\sim 5\times 10^7\,\Mo$,
well below the masses expected for the haloes of luminous galaxies but
far above the lower limit for subhaloes in a \LCDM\ universe which
could be as low as
$10^{-12}\,\Mo$~\citep{Hofmann2001,Bertone2005}. Considerable
extrapolation is thus necessary in order to estimate the total subhalo
emission. Note that even at the Ph-A-1 resolution threshold of
$5\times 10^7\,\Mo$, the surface brightness is dominated by the
subhalo component at radii greater than $200\,{\rm kpc}$.

To calibrate the extrapolation to lower subhalo masses we combine
results from our nine Phoenix simulations with results from six higher
resolution simulations of galaxy haloes from the Aquarius
Project~\citep{sp08a,sp08b}. Figure~2 shows the total annihilation
luminosity per unit halo mass ($M_{200}$) and per decade in subhalo
mass from subhaloes with masses ranging over 7 orders of magnitude,
from $10^5$ to $10^{12}\,\Mo$. In the overlap region between $10^8$
and $10^9\,\Mo$, the Phoenix and Aquarius results agree to about
30\%. This is within the scatter expected given the finite number of
realizations (illustrated by the shaded area) and the roll-off as
subhalo mass approaches 1\% of the parent mass. Well away from these
cutoffs, the shape of this curve is very similar to that of the halo
luminosity per unit mass expected for the Universe as a whole, shown as
the dashed magenta curve in Figure~2. This reflects the fact that the
luminosity is dominated by subhaloes in the outer regions which were
accreted recently~\citep{Gao2004} and so have similar luminosities and
abundance per unit mass (apart from a small bias correction of 1.5) as
the haloes in a representative volume of the Universe. Thus, we can
use analytic predictions for the abundance and concentration of field
haloes~\citep{st,Neto2007} to extrapolate our simulation
results to much lower subhalo masses. The upper blue curves in
Figure~1 show the resulting predictions for minimum subhalo masses of
$10^{-6}$ and $10^{-12}\,\Mo$, respectively. The most uncertain part
of this extrapolation is the assumption that halo concentration
continues to increase towards lower masses in the same way as measured
over the mass range simulated so far. This assumption has not yet
tested explicitly, and has a very large effect on the results. For
example, if all (sub)haloes less massive than $10^5\,\Mo$ are assumed
to have similar concentration, then the total predicted emission from
subhaloes would be more than two orders of magnitude below that
plotted in Figure~1 for an assumed cut-off mass of $10^{-6}\,\Mo$.

\begin{figure}
\begin{center}
\resizebox{8cm}{!}{\includegraphics{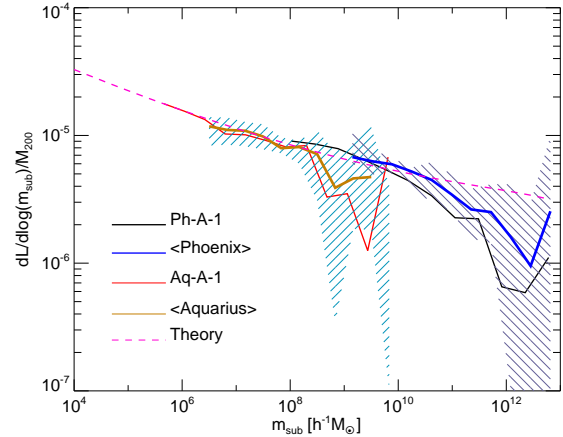}}
\end{center}
\caption{Annihilation luminosity (in arbitrary units) from subhaloes lying within $r_{200}$
  per decade in subhalo mass and per unit halo mass ($M_{200}$) for
  the Phoenix and Aquarius simulations. The level-1 simulations are
  shown by the black (Phoenix) and red (Aquarius) lines
  and the medians of the nine Phoenix and six Aquarius level-2
  simulations by the thick blue and orange lines respectively. The full
  scatter in each set of simulations is indicated by the shaded
  areas. The dashed magenta line gives the predicted annihilation
  luminosity density per decade in halo mass from the cosmic
  population of dark matter haloes. }
\label{scatter}
\end{figure}

With our adopted concentration scaling, subhaloes dominate the surface
brightness beyond projected radii of a few kiloparsecs, as may be
seen in Fig.~1. Surface brightness is almost constant between 10 and
$300\,{\rm kpc}$, dropping by a factor of two only at $460\,{\rm
kpc}$. At the virial radius of the cluster ($r_{200}=1936~{\rm kpc}$),
the surface brightness of the subhalo component is a factor of 14
below its central value. Within this radius the luminosity from
resolved subhaloes in Ph-A-1 is more than twice that from the smooth
halo, even though these subhaloes account only for 8\% of the
mass. Extrapolating to minimum subhalo masses of $10^{-6}$ and
$10^{-12}\,\Mo$ the subhalo excess becomes $718$ and $16089$
respectively.  These boost factors substantially exceed the equivalent
factors predicted for the galaxy haloes of the Aquarius Project. 
This is because of the additional high-mass subhaloes which contribute
in the cluster case (see Figure~2) together with the lower
concentration of cluster haloes relative to galaxy haloes, which
reduces the emission from the smooth component. Note, the boost
factor for the Aq-A-1 obtained with the extrapolation we use here is
smaller by a factor of 2.4 than the value quoted in Springel et
al. (2008a).

For the resolved component, there is significant variation amongst the
nine Phoenix haloes, but the median value of the total boost factor
(for a cutoff mass of $10^{-6}M_{\odot}$) is 1125, which, for the
reasons just given, is about twelve times the median boost factor we
obtain by applying the same method to the Aquarius haloes. 
Comparing these results suggests that the ratio of
subhalo to smooth main halo luminosity within $r_{200}$ (subhalo
``boost factor'') varies with halo mass approximately as

\begin{equation}
b(M_{200}) = L_{\rm sub}/L_{\rm main} = 1.6 \times 10^{-3}(M_{200}/{\rm M_\odot})^{0.39}.
\end{equation}
The total luminosity of a halo is therefore $L_{\rm tot} = (1 +
b)L_{\rm main}$, where $L_{\rm main}$ is the emission of the smooth
halo. In addition, the projected luminosity profile of the subhalo
component can be well approximated by
\begin{equation}
S_{\rm sub}(r) = \frac{16 b(M_{200})L_{\rm main}}{\pi\ln (17)}\frac{1}{r_{200}^2+16r^2}.
\end{equation}
These formulae will be used to estimate dark matter annihilation
luminosities and surface brightness profiles for haloes with different
masses in subsequent sections.

\subsection{Surface brightness and signal to noise of galaxies and clusters}

\begin{figure*}
\begin{center}
\resizebox{8.0cm}{!}{\includegraphics{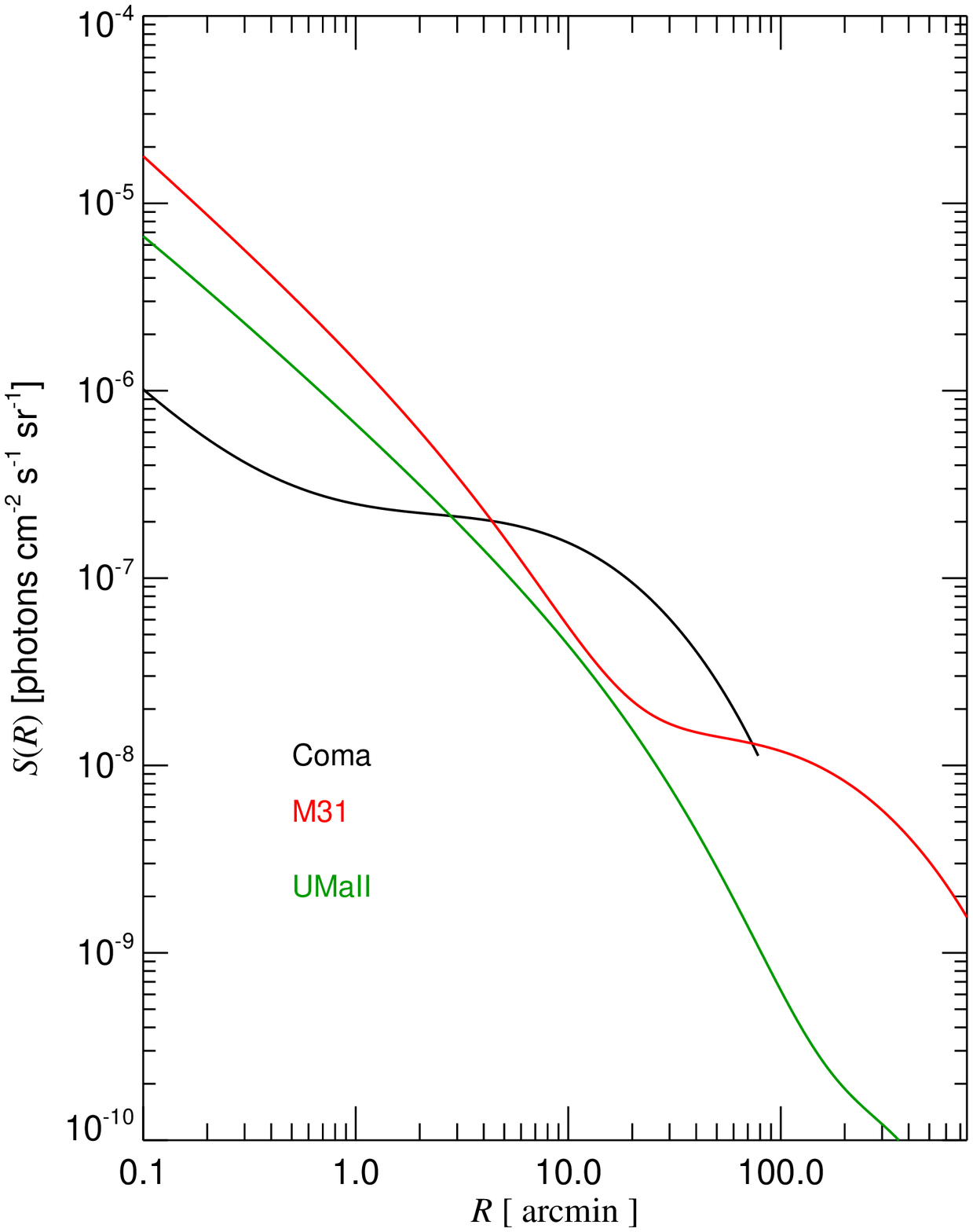}}%
\resizebox{8.0cm}{!}{\includegraphics{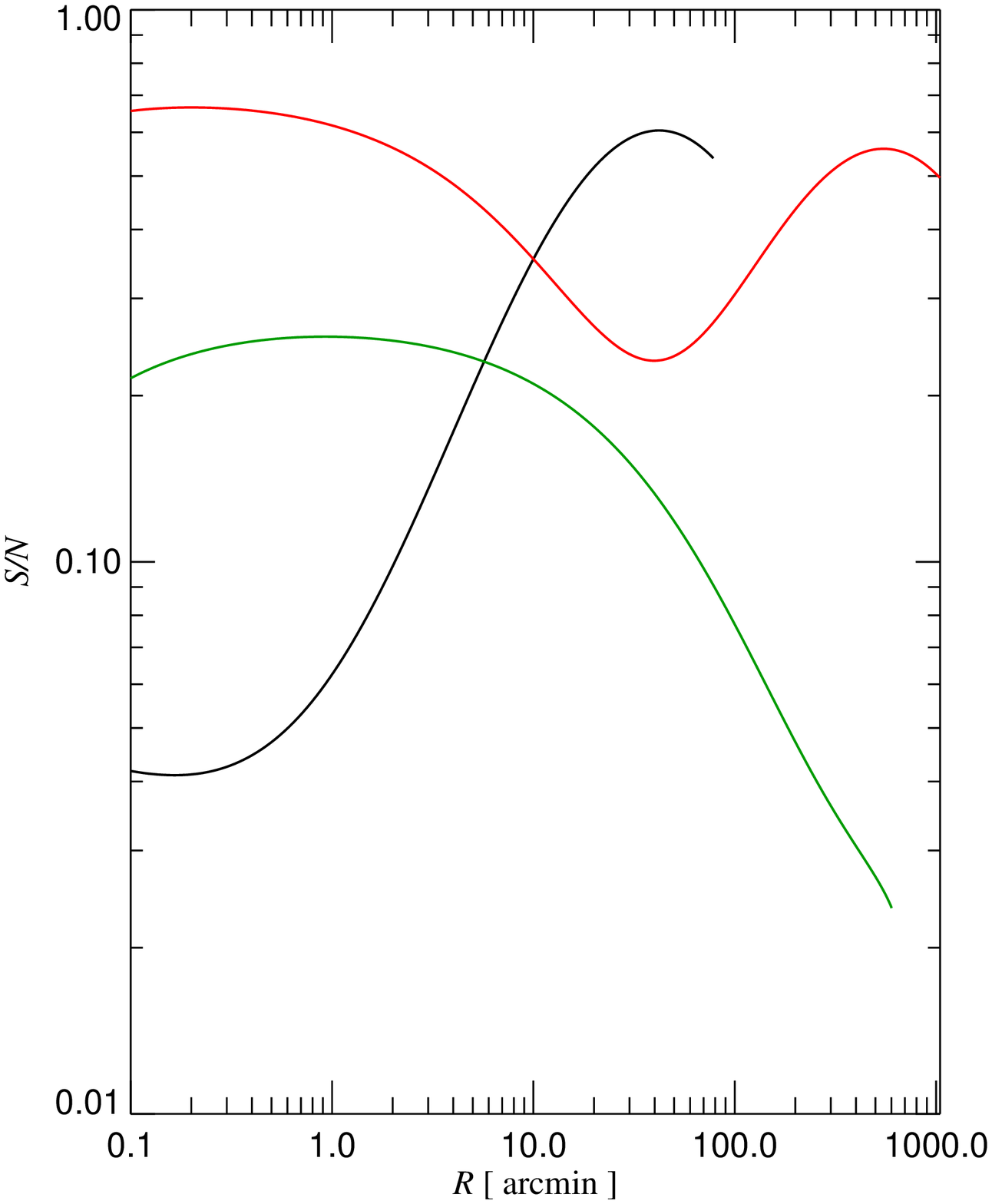}}%
\caption{{\em Left panel:} Predicted surface brightness profiles of
  annihilation radiation (in units of annihilation photons per ${\rm
    cm}^2$ per second per steradian) for a dwarf galaxy (UMaII; green
  line), for the nearest large galaxy (M31; red line) and for a rich
  galaxy cluster (the Coma cluster; black line). As in Figure~1,
  surface brightness scales as $N_{\gamma} \langle \sigma
  v\rangle/m_p^2$. Projected radius is given in arc minutes. The inner
  steeply rising part of each curve is due to smoothly distributed
  dark matter in the main halo, while the shoulder of extended
  emission is produced by low-mass subhaloes. Each profile is
  truncated at $r_{200}$, the nominal radius of the dark matter
  halo. {\em Right panel:} Estimates of the signal-to-noise ratio
  within a circular aperture of radius $R$ (in arc minutes). The
  signal is obtained by direct integration of the corresponding curves
  in the left-hand panel and the noise is obtained as discussed in the
  text. $S/N$ scales as $N_{\gamma}\,B^{-1/2} \langle \sigma
  v\rangle/m_p^2$, where $B$ is the surface brightness of the
  background, assumed to be uniform.}
\end{center}
\end{figure*}

Putting together results from the Phoenix and Aquarius projects, we
can assess the relative ease of detection of cluster, galaxy and dwarf
satellite haloes. In the left panel of Figure~3, we show predicted
surface brightness profiles for three of the most promising
candidates, the Coma cluster of galaxies, the Andromeda Nebula (M31)
and the dwarf satellite galaxy, Ursa Major-II (UMa-II), assuming a
minumum subhalo mass of $10^{-6}\,\Mo$. We represent Coma and M31 by
scaling Ph-A-1 and Aq-A-1 to the appropriate virial masses,
$M_{200}=1.3\times 10^{15}\,\Mo$ for Coma~\citep{reiprich02} and $1.8 \times
10^{12}\,\Mo$ for M31~\citep{lw08}.  We model UMa-II as
in~\cite{sp08a} (including the contribution from substructures -- the
`subsub' component). At projected radii below 2 arcmin, M31 is about
twice as bright as UMa-II and both are substantially brighter than
Coma. However, at 20 arcmin the surface brightness of Coma exceeds
that of M31 by a factor of 4 and that of UMa-II by about a factor of
6. Beyond about 70 arcmins, M31 is again the brightest object.

For $\gamma$-ray telescopes like the Fermi Large Area Telescope
(LAT), the detectability of extended objects depends on their
contrast relative to the diffuse background. As a simple indicator of
signal-to-noise ($S/N$), in the right panel of Figure~3 we estimate
the signal within a circular aperture from the enclosed luminosity,
and the noise as the square root of the background counts, assumed to
be $B\times A \times t$, where $B$ is the background count rate per
unit area, $A$ is the area of the aperture in square arc minutes, and
$t$ is the exposure time. (This assumes that the background is uniform
and larger than the signal, which may not be the case for the smallest
apertures.) For the dwarf galaxy UMa-II, the effective $S/N$ is almost
independent of aperture for radii less than 10 arcmin, but drops
dramatically at larger radii. In contrast, the $S/N$ for Coma rises
steeply with increasing aperture to a peak at a radius of about 30
arcmin, significantly larger than the few arcmin resolution of the
Fermi-LAT at energies $\sim 10$~GeV. For M31, the effective $S/N$
has a minimum on this scale and has maxima on scales of one and
300~arcmin. In this simple set-up the maximum achievable $S/N$ ratios
for Coma and M31 exceed that for UMa-II by about a factor of 3.

In practice, realistic experiments will find it difficult to achieve
these theoretical $S/N$ values for very large apertures. Systematic
effects due to variable backgrounds and difficulties in masking bright
sources make background correction significantly easier for small
apertures.  M31 is a particularly difficult case because of its very
large angular size, its low galactic latitude, and confusion from
other $\gamma$-ray sources in its inner regions. The Coma cluster is
significantly more promising because it lies close to the North
Galactic Pole and appears 10 times smaller on the sky. On the other
hand, an overly small aperture, corresponding for example to the few
arcmin resolution of the Fermi LAT instrument at about $10\,{\rm
  GeV}$, would miss a large fraction of the signal in Coma and other
nearby galaxy clusters. For a uniform background, the optimal filter
has a shape similar to the predicted profile~\citep{sp08a}
shown in Figure~3 and represented by equation (2).

\section{Discussion and conclusion}
\begin{table*}
\begin{center}
\begin{tabular}{lcccccc}
\hline
\hline
Object Name & Half-light radius & Distance & $M_{\rm 200}$  & $L$ &$F=L/(4\pi d^2)$ & $S/N$ \\

&[arcmin] &[Mpc] &[${\rm M_\odot}$] &[$L_{\rm mw}$] &[$F_{\rm mw}$]
&[$(S/N)_{\rm mw}$]\\
\hline
\hline
AWM 7  & 35.5 & 67.0 & $4.2 \times 10^{14}$ &$7.1 \times 10^4$ & $3.2 \times 10^{-4}$
& $6.8 \times 10^{-3}$ \\
Fornax Cluster & 84.1 & 17.5 & $1.0 \times 10^{14}$ &$1.2 \times 10^4$
& $8.0 
\times 10^{-4}$ & $7.3 \times 10^{-3} $\\
M49    & 59.6 & 18.2 &$0.4 \times 10^{14}$ &$3.9 \times 10^3$ &$2.4 \times 10^{-4}$ &$3.1 \times 10^{-3}$ \\
NGC 4636 &52.6 &  17.4 & $0.24 \times 10^{14}$ &$2.1 \times 10^3$
&$1.4 \times 10^{-4}$ &$2.0 \times 10^{-3}$ \\
Centaurus (A3526) &40.1  & 50.5 & $2.6 \times 10^{14}$ &$3.9 \times
10^4$ &$3.1 \times 10^{-4}$ &$5.8 \times 10^{-3}$ \\
Coma & 36.1 & 95.8 &$1.3 \times 10^{15}$ &$2.9 \times 10^5$ &$6.4
\times 10^{-4}$ &$1.3 \times 10^{-2}$ \\
\hline
\hline
Draco &16.4 &0.082 & N/A &$5.2 \times 10^{-3}$ &$1.6 \times 10^{-5}$
&$6.3 \times 10^{-4}$ \\
UMaI  &18.4  &0.066 & N/A &$4.3 \times 10^{-3}$ &$2.0 \times 10^{-5}$
& $7.5 \times 10^{-4}$ \\
LeoI &4.4 &0.25 & N/A &$3.5 \times 10^{-3}$ & $1.2 \times 10^{-6}$
&$8.2 \times 10^{-5}$ \\
Fornax dwarf &5.9 & 0.138 & N/A &$2.0 \times 10^{-3}$ &$2.2 \times
10^{-6}$ &$1.5 \times 10^{-4}$ \\
LeoII &2.5 & 0.205 & N/A &$8.5 \times 10^{-4}$ &$4.1 \times 10^{-7}$
&$3.1 \times 10^{-5}$ \\
Carina &4.6 & 0.101 & N/A &$7.1 \times 10^{-4}$ &$1.4 \times 10^{-6}$
&$1.0 \times 10^{-4}$ \\
Sculpt &13.2 & 0.079 & N/A &$3.2 \times 10^{-3}$ &$1.0 \times 10^{-5}$
&$4.9 \times 10^{-4}$ \\
Sext &3.3 & 0.086 & N/A &$3.0 \times 10^{-4}$ &$8.3 \times 10^{-7}$
&$6.1 \times 10^{-5}$ \\
UMaII &28.8 &0.032 & N/A &$2.6 \times 10^{-3}$ &$5.2 \times 10^{-5}$
&$1.3 \times 10^{-3}$ \\
Comber &15.9 &0.044 & N/A &$1.6 \times 10^{-4}$ &$1.7 \times 10^{-5}$
&$6.8 \times 10^{-4}$ \\
WilI &17.7 &0.066 & N/A &$3.9 \times 10^{-3}$ &$1.8 \times 10^{-5}$
&$7.0 \times 10^{-4}$ \\
LMC &82.5 &0.049 & N/A &$3.8 \times 10^{-2}$ &$3.3 \times 10^{-4}$
&$3.1 \times 10^{-3}$ \\
SMC &45.5 &0.061 & N/A &$1.9 \times 10^{-2}$ &$1.1 \times 10^{-4}$
&$1.8 \times 10^{-3}$ \\
\hline
\hline
M31 &351.5 &0.807 &$1.8 \times 10^{12}$ &$1.3 \times 10^{2}$ &$4.2
\times 10^{-3}$ &$9.3 \times 10^{-3}$ \\ 
\hline
\hline
\end{tabular}
\end{center}
\caption{Principal properties of nearby galaxy clusters, prominent
  satellites of the Milky Way, and the Andromeda Nebula, M31. The
  annihilation luminosity, $L$, is given in units of the luminosity
  from the smooth component of the main Aq-A halo, which we use as a
  proxy for the Milky Way.  The observed flux, $F$, is expressed
  relative to the flux received by an observer placed $8\,{\rm kpc}$
  from the centre of Aq-A. Similarly, the predicted $S/N$ for an
  optimal filter placed on each object is normalized to the
  signal-to-noise predicted for a similar filter tuned to the diffuse
  emission of Aq-A seen from this observer location. For the
  signal-to-noise calculations, we use the optimal filter
  of~Springel et al. (2008a) assuming the background to be the same
  everywhere and to dominate the signal in all objects.}
\label{TabObes}
\end{table*}

In Table~\ref{TabObes} we summarise properties of some nearby
astronomical objects which are relevant for the detectability of their
dark matter annihilation signal.  We consider six galaxy clusters
which were already analyzed by the Fermi collaboration
\citep{ackermann10}, thirteen of the known dwarf satellites of our
Galaxy, and the nearest giant galaxy, M31.

For the galaxy clusters, distances were taken from the
NASA/IPAC Extragalactic Database\footnote{http://nedwww.jpac.caltech.edu/},
and virial masses, ${\rm M_{200}}$ (based on X-ray data), from
\cite{reiprich02}. Values for $V_{\rm max}$ and $r_{\rm max}$
were derived assuming an NFW density profile~\citep{NFW96,NFW97} and the
mass-concentration relation of \cite{Neto2007}. We have verified that
this relation is consistent with our simulation data down to the
resolution limit of Aq-A-1, which is about $10^5\, {\rm M_{\odot}}$.

Data for dwarf satellites were taken from the mass models of
\cite{penarrubia08}. Their $\gamma$-ray luminosities are estimated
from an emission integral based on the NFW formula, $\int \rho^2 dV =
1.23\,V_{\rm max}^4/(G^2r_{\rm max})$. As discussed in
\cite{sp08a}, the annihilation signal due to substructures
within Milky Way dwarfs (the `subsub' component) is less than that due
to the smooth component of their haloes in almost all cases, so we do
not consider it here. The distance of M31 was also taken from the
NASA/IPAC Extragalactic Database. We base structural parameters for
the M31 halo on the Aq-A-1 simulation which has a very similar
mass~\citep{lw08}.

We estimate a ``best case'' signal-to-noise for each object using the
optimal filter discussed by \cite{sp08a} and assuming a uniform
background accross the whole sky which dominates over the signal in
all objects.  In this case, the optimal filter has the same shape as
the signal, and the signal-to-noise can be written in the generic form
\begin{equation}
S/N = f_{\rm shape}(\theta_h/\theta_{\rm psf})\,\left[\frac{t A_{\rm eff}}{B}\right]^{1/2} \frac{F}{(\theta^2_{h}+\theta^2_{\rm psf})^{1/2}},
\end{equation}
where $F = L/(4\pi d^2)$ is the photon flux, $\theta_h$ the half-light
radius, $\theta_{\rm psf}$ ($\simeq 10$ arcmin for Fermi at the
relevant energies~\citep{Michelson07}) describes the point spread
function of the instrument, $t$ is the integration time, $A_{\rm eff}$
is the effective collecting area of the telescope, and $B$ is the
background count rate per unit solid angle. The function $f_{\rm
shape}(x)$ encodes the detailed shape of the emission profile of the
signal~\citep{sp08a}; it is of order unity and depends only weakly on
the ratio $x=\theta_h/\theta_{\rm psf}$.

Using the techniques discussed above, we can compare the apparent
$\gamma$-ray luminosities and achievable $S/N$ ratios for galaxy
clusters with those estimated by \cite{sp08a} for dwarf satellites of
the Milky Way. Results are shown in the Table. We find that the
brightest nearby cluster, Fornax, is predicted to appear 15 times more
luminous than the brightest dwarf spheroidal, UMaII, and 40-50 times
more luminous than UMaI, Draco or the ultrafaint satellite, Wilman-1.
However, the Fornax cluster is quite extended on the sky, and, as a
result, when optimal filters are used, the slightly fainter but more
compact Coma cluster has a predicted $S/N$ ratio 1.8 times larger and
ten times that of the most easily detectable dwarf spheroidal,
UMaII. Although the Andromeda Nebula is predicted to have comparable
$S/N$, it is not a promising target because of the difficulty in
correcting for foreground and other sources of emission. Note that the
$S/N$ predicted for both objects is still very small compared to that
of the main component of the Milky Way's smooth halo. Here also, of
course, the main problem is in separating annihilation radiation from
other $\gamma$-ray signals.

The Coma cluster thus offers an order of magnitude better opportunity
than any Milky Way satellite for detecting dark matter or placing
limits on its annihilation cross-section. As we have shown, for a high
resolution experiment like Fermi, the sensitivity for detecting such
radiation will be enhanced by use of a filter which is properly
matched to the expected extent of the object. For example, for the
optimal filter, the $S/N$ expected for Coma is about 1.5 times higher
than the $S/N$ for a filter based on the point-spread function of the
Fermi LAT, assuming 10 arcmin for the latter at the relevant energies.
Detecting annihilation radiation from the Coma or Fornax clusters or
the placing of robust and stringent upper limits will also require
careful subtraction of astrophysical sources and an accurate estimate
of the background.

\section*{acknowledgements}
We thank the Supercomputer Center of the Chinese Academy of Science,
where the simulations were carried out. LG acknowledges support from
the one-hundred-talents program of the Chinese academy of
science (CAS), the National basic research program of China (973
program under grant No. 2009CB24901), the {\small NSFC} grants program
(No. 10973018), the Partner Group program of the Max Planck Society,
and an STFC Advanced Fellowship, as well as the hospitality of the
Institute for Computational Cosmology in Durham, UK. CSF acknowledges
a Royal Society Wolfson Research Merit Award and ERC Advanced
Investigator grant COSMIWAY. This work was supported in part by an
STFC rolling grant to the ICC.
\bibliographystyle{mnras}
\bibliography{dm}

\begin{thebibliography}{36}
\expandafter\ifx\csname natexlab\endcsname\relax\def\natexlab#1{#1}\fi

\bibitem[{Abdo} et~al.(2010){Abdo}, {Ackermann}, {Ajello} et~al.]{abdo}
{Abdo} A.~A., {Ackermann} M., {Ajello} M., et~al., 2010, \apj, 712, 147

\bibitem[{Ackermann} et~al.(2010){Ackermann}, {Ajello}, {Allafort}
  et~al.]{ackermann10}
{Ackermann} M., {Ajello} M., {Allafort} A., et~al., 2010, \jcap, 5, 25

\bibitem[{Anderson} et~al.(2010){Anderson}, {Kuhlen}, {Diemand}, {Johnson} \&
  {Madau}]{Anderson2010}
{Anderson} B., {Kuhlen} M., {Diemand} J., {Johnson} R.~P., {Madau} P., 2010,
  \apj, 718, 899

\bibitem[{Berezinsky} et~al.(1994){Berezinsky}, {Bottino} \&
  {Mignola}]{Berezinsky1994}
{Berezinsky} V., {Bottino} A., {Mignola} G., 1994, Physics Letters B, 325, 136

\bibitem[{Berezinsky} et~al.(2003){Berezinsky}, {Dokuchaev} \&
  {Eroshenko}]{Berezinsky2003}
{Berezinsky} V., {Dokuchaev} V., {Eroshenko} Y., 2003, \prd, 68, 10, 103003

\bibitem[{Bergstr{\"o}m} et~al.(1998){Bergstr{\"o}m}, {Ullio} \&
  {Buckley}]{Bergstrom1998}
{Bergstr{\"o}m} L., {Ullio} P., {Buckley} J.~H., 1998, Astroparticle Physics,
  9, 137

\bibitem[{Bertone} et~al.(2005){Bertone}, {Hooper} \& {Silk}]{Bertone2005}
{Bertone} G., {Hooper} D., {Silk} J., 2005, \physrep, 405, 279

\bibitem[{Colafrancesco} et~al.(2007){Colafrancesco}, {Profumo} \&
  {Ullio}]{Colafrancesco2007}
{Colafrancesco} S., {Profumo} S., {Ullio} P., 2007, \prd, 75, 2, 023513

\bibitem[{Diemand} et~al.(2007){Diemand}, {Kuhlen} \& {Madau}]{Diemand2007}
{Diemand} J., {Kuhlen} M., {Madau} P., 2007, \apj, 657, 262

\bibitem[{Diemand} et~al.(2008){Diemand}, {Kuhlen}, {Madau} et~al.]{diemand08}
{Diemand} J., {Kuhlen} M., {Madau} P., et~al., 2008, \nat, 454, 735

\bibitem[{Gao} et~al.(2004){Gao}, {White}, {Jenkins}, {Stoehr} \&
  {Springel}]{Gao2004}
{Gao} L., {White} S.~D.~M., {Jenkins} A., {Stoehr} F., {Springel} V., 2004,
  \mnras, 355, 819

\bibitem[{Green} et~al.(2005){Green}, {Hofmann} \& {Schwarz}]{Green05}
{Green} A.~M., {Hofmann} S., {Schwarz} D.~J., 2005, \jcap, 8, 3

\bibitem[{Hofmann} et~al.(2001){Hofmann}, {Schwarz} \&
  {St{\"o}cker}]{Hofmann2001}
{Hofmann} S., {Schwarz} D.~J., {St{\"o}cker} H., 2001, \prd, 64, 8, 083507

\bibitem[{Jeltema} et~al.(2009){Jeltema}, {Kehayias} \& {Profumo}]{jeltema10}
{Jeltema} T.~E., {Kehayias} J., {Profumo} S., 2009, \prd, 80, 2, 023005

\bibitem[{Kamionkowski} et~al.(2010){Kamionkowski}, {Koushiappas} \&
  {Kuhlen}]{Kamionkowski2010}
{Kamionkowski} M., {Koushiappas} S.~M., {Kuhlen} M., 2010, \prd, 81, 4, 043532

\bibitem[{Komatsu} et~al.(2011){Komatsu}, {Smith}, {Dunkley} et~al.]{Komatsu11}
{Komatsu} E., {Smith} K.~M., {Dunkley} J., et~al., 2011, \apjs, 192, 18

\bibitem[{Koushiappas} et~al.(2004){Koushiappas}, {Zentner} \&
  {Walker}]{Koushiappas2004}
{Koushiappas} S.~M., {Zentner} A.~R., {Walker} T.~P., 2004, \prd, 69, 4, 043501

\bibitem[{Kuhlen} et~al.(2008){Kuhlen}, {Diemand} \& {Madau}]{Kuhlen2008}
{Kuhlen} M., {Diemand} J., {Madau} P., 2008, \apj, 686, 262

\bibitem[{Li} \& {White}(2008)]{lw08}
{Li} Y.-S., {White} S.~D.~M., 2008, \mnras, 384, 1459

\bibitem[{Michelson}(2007)]{Michelson07}
{Michelson} P.~F., 2007, in { The First GLAST Symposium\/}, edited by {S.~Ritz,
  P.~Michelson, \& C.~A.~Meegan}, vol. 921 of { American Institute of Physics
  Conference Series\/},  8--12

\bibitem[{Navarro} et~al.(1996){Navarro}, {Frenk} \& {White}]{NFW96}
{Navarro} J.~F., {Frenk} C.~S., {White} S.~D.~M., 1996, \apj, 462, 563

\bibitem[{Navarro} et~al.(1997){Navarro}, {Frenk} \& {White}]{NFW97}
{Navarro} J.~F., {Frenk} C.~S., {White} S.~D.~M., 1997, \apj, 490, 493

\bibitem[{Neto} et~al.(2007){Neto}, {Gao}, {Bett} et~al.]{Neto2007}
{Neto} A.~F., {Gao} L., {Bett} P., et~al., 2007, \mnras, 381, 1450

\bibitem[{Pe{\~n}arrubia} et~al.(2008){Pe{\~n}arrubia}, {Navarro} \&
  {McConnachie}]{penarrubia08}
{Pe{\~n}arrubia} J., {Navarro} J.~F., {McConnachie} A.~W., 2008, \apj, 673, 226

\bibitem[{Pieri} et~al.(2008){Pieri}, {Bertone} \& {Branchini}]{Pieri2008}
{Pieri} L., {Bertone} G., {Branchini} E., 2008, \mnras, 384, 1627

\bibitem[{Pinzke} et~al.(2011){Pinzke}, {Pfrommer} \& {Bergstrom}]{Pinzke11}
{Pinzke} A., {Pfrommer} C., {Bergstrom} L., 2011, ArXiv:1105.3240

\bibitem[{Reiprich} \& {B{\"o}hringer}(2002)]{reiprich02}
{Reiprich} T.~H., {B{\"o}hringer} H., 2002, \apj, 567, 716

\bibitem[{Sanchez-Conde} et~al.(2011){Sanchez-Conde}, {Cannoni}, {Zandanel},
  {Gomez} \& {Prada}]{sanchez}
{Sanchez-Conde} M.~A., {Cannoni} M., {Zandanel} F., {Gomez} M.~E., {Prada} F.,
  2011, ArXiv e-prints
\bibitem[{Sheth} \& {Tormen}(2002)]{st}
{Sheth} R.~K., {Tormen} G., 2002, \mnras, 329, 61

\bibitem[{Spergel} et~al.(2003){Spergel}, {Verde}, {Peiris} et~al.]{spergel03}
{Spergel} D.~N., {Verde} L., {Peiris} H.~V., et~al., 2003, \apjs, 148, 175

\bibitem[{Springel} et~al.(2008{\natexlab{a}}){Springel}, {White}, {Frenk}
  et~al.]{sp08a}
{Springel} V., {White} S.~D.~M., {Frenk} C.~S., et~al., 2008{\natexlab{a}},
  \nat, 456, 73

\bibitem[{Springel} et~al.(2008{\natexlab{b}}){Springel}, {Wang},
  {Vogelsberger} et~al.]{sp08b}
{Springel} V., {Wang} J., {Vogelsberger} M., et~al., 2008{\natexlab{b}},
  \mnras, 391, 1685


\bibitem[{Springel} et~al.(2005){Springel}, {White}, {Jenkins} et~al.]{sp05}
{Springel} V., {White} S.~D.~M., {Jenkins} A., et~al., 2005, \nat, 435, 629

\bibitem[{Stoehr} et~al.(2003){Stoehr}, {White}, {Springel}, {Tormen} \&
  {Yoshida}]{Stoehr2003}
{Stoehr} F., {White} S.~D.~M., {Springel} V., {Tormen} G., {Yoshida} N., 2003,
  \mnras, 345, 1313

\bibitem[{Strigari} et~al.(2008){Strigari}, {Koushiappas}, {Bullock}
  et~al.]{Strigari2008}
{Strigari} L.~E., {Koushiappas} S.~M., {Bullock} J.~S., et~al., 2008, \apj,
  678, 614

\bibitem[{Vogelsberger} \& {White}(2011)]{Vogelsberger11}
{Vogelsberger} M., {White} S.~D.~M., 2011, \mnras, 413, 1419

\bibitem[{Zavala} et~al.(2010){Zavala}, {Springel} \&
  {Boylan-Kolchin}]{Zavala2010}
{Zavala} J., {Springel} V., {Boylan-Kolchin} M., 2010, \mnras, 405, 593

\end{thebibliography}
\end{document}